\documentclass{ws-book9x6}
\usepackage{ws-book-thm}   
\usepackage{ws-book-har}   
\usepackage[pdfpagelabels=false,colorlinks=true,allcolors=black]{hyperref}  

\newcommand\summaryname{Abstract}
    {\small\begin{center}%
    \bfseries{\summaryname} \end{center}}

\definecolor{ashgrey}{rgb}{0.52, 0.52, 0.51}

\newcommand{\bx}{\mathbf{x}}
\newcommand{\bv}{\mathbf{v}}
\newcommand{\bw}{\mathbf{w}}

\newcommand{\R}{\mathbb{R}}
\newcommand{\E}{\mathbb{E}}
\newcommand{\N}{\mathcal{N}}

\newcommand{\dd}{\mathrm{d}}

\newcommand{\dimx}{N}
\newcommand{\nhid}{K}
\newcommand{\ndata}{P}
\newcommand{\dataset}{\mathcal{D}}
\DeclareMathOperator{\sign}{sign}

\newcommand{\mb}{\mathbf}
\newcommand{\Wl}[1]{{W}^{#1}}

\newcommand{\x}{\mathbf{x}}
\newcommand{\h}{\mathbf{h}}
\newcommand{\bi}{\mathbf{b}}

\title{Replica Symmetry Breaking \& Far Beyond}
\makeindex
\allowdisplaybreaks
\begin{document}

\renewcommand{\thechapter}{24}
\chapter[Neural networks: from the perceptron to deep nets]{Neural networks: from the perceptron to deep nets}\label{ch24}

\author{Marylou Gabrié}
\address{
\textsuperscript{$*$}Department of Computing Sciences, Bocconi University, Milano, Italy\\
\textsuperscript{$\dagger$}Bocconi Institute for Data Science and Analytics, Milano, Italy\\
}

\author{Surya Ganguli}
\address{
Department of Applied Physics, Stanford University, Stanford, USA}
\author{Carlo Lucibello and Riccardo Zecchina}
\address{
Department of Computing Sciences, Bocconi University, Milano, Italy\\
Bocconi Institute for Data Science and Analytics, Milano, Italy\\
}

\vspace{1cm}

Artificial networks have been studied through the prism of statistical mechanics as disordered systems since the 80s, starting from the simple models of Hopfield’s associative memory and the single-neuron perceptron classifier. Assuming data is generated by a teacher model, asymptotic generalisation predictions were originally derived using the replica method and the online learning dynamics has been described in the large system limit. In this chapter, we review the key original ideas of this literature along with their heritage in the ongoing quest to understand the efficiency of modern deep learning algorithms. One goal of current and future research is to characterize the bias of the learning algorithms toward well-generalising minima in a complex overparametrized loss landscapes with many solutions perfectly interpolating the training data. Works on perceptrons, two-layer committee machines and kernel-like learning machines shed light on these benefits of overparametrization. Another goal is to understand the advantage of depth while models now commonly feature tens or hundreds of layers. If replica computations apparently fall short in describing general deep neural networks learning, studies of simplified linear or untrained models, as well as the derivation of scaling laws provide the first elements of answers.

\section{Statistical physics approaches to learning problems}
\label{sec:perceptron}


The replica method  of spin glass theory has been successfully applied since the 80s to characterize the computational capacity of simple neural network models, starting with Hopfield's associative memory model and the simplest of neural classifiers, the perceptron. 
From the perspective of spin glass physics, the quenched disorder is given by the data, the patterns to be stored or classified. In the theoretical analysis of supervised learning scenarios, the data are often produced by a generative process, the teacher, of which the learning system, the student, may have prior statistical information. These studies concerned mainly  the generalization error, that is, the expected error produced by a trained network when a previously unseen data item is presented.  The results are derived in the asymptotic regime, in which the number of data points, the dimension of the input and the number of parameters are sent to infinity while maintaining a sensible scaling relationship between them. 
The statistical mechanics  studies of the 80s and 90s have lead to important advances, providing results on the storage and generalization capacity of nontrivial systems that were out of reach for mathematically rigorous techniques.

As far as the dynamics of learning processes is concerned, the contribution were mainly limited to the so called online setting, where the patterns are presented only once and the learning dynamics can  be described in the continuous time limit by a set of differential equations. When multiple passes over the dataset are involved instead, gradient descent and stochastic gradient descent on the perceptron model have been analyzed  using the much more complicated set of equations given by dynamical mean field theory.

In the  conceptual framework of the 1980s-1990s, the role of replica symmetry breaking was key. It allowed  to establish the limit of the learning capacity of non-convex classifiers and shed light on the geometrical structure of the loss landscape.

On the algorithmic side, the cavity method has led to the design of an efficient message-passing algorithm that can be used to obtain Bayesian or maximum likelihood predictions in simple neural architectures for which the method is exact (perceptrons, tree-like networks). The fixed points of the algorithm are related to the stationary points of the replica free energy, and the dynamics can be tracked statistically with a simple set of scalar equations called state evolution.
More complicated, and of much greater generality and relevance, is the analysis of algorithms such as gradient descent and stochastic gradient descent in non-convex landscapes and their link with good generalization. This is a complex challenge which requires understanding the interplay between complex algorithmic dynamics and the geometry of the learning landscape. The out-of-equilibrium dynamics of algorithms that implement learning processes represents an open conceptual challenge that is only minimally understood  and that appears to be essential for a thorough understanding of contemporary neural systems. Effective algorithms do not uniformly sample the solution space but seem to be biased toward configurations that generalized well.


\subsection{The storage problem}
The modern application of statistical physics to artificial neural networks had its origins in 1982, with the seminal introduction of the Hopfield model (HM)  \cite{Hopfield1982}. The HM was created as a toy biological model of an associative memory, whose goal is to store $\ndata$ binary configurations, called \emph{memories} or \emph{patterns}, that represent the firing or non-firing state of $N$ neurons. 
The prescribed memories $\bx^\mu \in \{-1,+1\}^N$, for $\mu=1,\dots,\ndata$, are by definition successfully stored if the neural network dynamics has a fixed point very close to each pattern. In particular, Hopfield modeled the neural network dynamics as the zero temperature, greedy MCMC dynamics of an Ising spin system:
$\sigma_i^{t+1}=\sign(\sum_{j\neq i} J_{ij} \sigma^t_j)$. The synaptic connectivity matrix $J$ is chosen to store the memories via the Hebb rule \cite{Hebb1949-th}: $J_{ij} = \sum_{\mu=1}^\ndata \xi^\mu_i\xi^\mu_j$. With a simple signal-to-noise argument, in Ref. \cite{Hopfield1982} it is argued that network can store up to $\ndata \approx 0.14\dimx$ random patterns, yielding an extensive memory capacity proportional to the number of neurons. 

Note that the Hopfield dynamics corresponds to minimizing the energy function $E(\sigma) = -\frac{1}{N}\sum_{i<j} \sum_{\mu=1}^\ndata\xi_i^\mu \xi_i^\mu \sigma_i\sigma_j$.  Amit, Gutfreund and Sompolinsky \cite{Amit1987-gs} then precisely characterized the phase diagram of the statistical mechanics system associated with this energy function using replica theory, within a replica symmetric (RS) ansatz.
In the temperature $T$ vs memory load $\alpha=\ndata/
\dimx$ plane, they found at low $\alpha$ and low $T$ a ferromagnetic region where the memories correspond to stable states that dominate the equilibrium Gibbs state. These states were called retrieval states. At higher $\alpha$ and low $T$, the retrieval states still exist as metastable states, but the equilibrium is dominated by mixtures of a finite number of patterns. Finally, for high $\alpha$ and low $T$ there is a pure spin glass phase, while at high $T$ there is a paramagnetic phase. In both cases the retrieval states are no longer present.  
Further analysis showing the existence of a replica symmetry breaking (RSB) instability and the application of the 1RSB formalism, only slightly refines the RS estimates \cite{crisanti_saturation_1986}.

The Hebb rule, while neurobiologically motivated, is however only one of many possible ways to store the $\ndata$ memories. This raises the natural question of what might be the highest possible storage capacity over all possible choices of connectivity $J$. The answer to this question, under the statistical assumption of random and independently generated memory patterns, was given by the seminal calculations of Gardner and Derrida for continuous synaptic strengths 
\cite{Gardner1988, Gardner1988a} and Mézard and Krauth for discrete ones \cite{Mezard1989}.  The foundational idea was to consider the space of all possible connectivity matrices $J$ that are consistent with the memory storage fixed point conditions $\xi^\mu_i=\sign(\sum_{j\neq i} J_{ij} \xi^\mu_j)$ for every memory $\mu$.  Each memory therefore imposes a constraint on $J$. Moreover, these constraints decouple, or are independent, across every row of $J$. Now a single row of $J$ corresponds to the incoming synaptic weights onto a single neuron. Thus the problem of calculating the storage capacity of an associative memory with pairwise interactions reduces to that of a single neuron (perceptron).

Let $\bw$ denote vector of synaptic weights onto any one neuron, corresponding to some row of $J$.
The approach pioneered by Gardner was to compute the volume of allowed synaptic weight configurations consistent with the storage of a dataset of $\ndata$ examples, $\dataset=\{\bx^\mu,y^\mu\}_{\mu=1}^\ndata$. This volume can be computed via the partition function
\begin{equation}
    Z_\dataset = \int \dd P(\bw)\ \prod_{\mu=1}^\ndata\Theta\left(\frac{1}{\sqrt{\dimx}}y^\mu\sum_{i=1}^\dimx w_i x^\mu_i\right).
\end{equation}
Here $\Theta(x)$ is the Heaviside function, $\Theta(x) = 1$ if $x>0$ and $0$ otherwise and $P(\bw)$ is the uniform measure on the set of allowed perceptron weights (for continuous weights this is the hypersphere  $\sum_i w^2_i = \dimx$, while for discrete weights this is the hypercube $\{-1,+1\}^\dimx$). 
We are interested in the {\it typical} volume in the high dimensional-limit $\ndata,\dimx \rightarrow \infty$ with finite $\alpha=\ndata/\dimx$. Therefore we consider the average \emph{entropy}
\begin{equation}
S(\alpha) = \lim_{\dimx\to\infty}\frac{1}{\dimx}\mathbb{E}_\dataset \log Z_\dataset.
\end{equation}
The expectation is over i.i.d. standard Gaussian inputs $x^\mu_i$ and over uniform i.i.d. $y^\mu\in\{-1,+1\}$ that are also independent of the inputs. As one requires storage of more patterns by increasing $\alpha$, the entropy and typical volume both decrease. Importantly, at a  critical capacity $\alpha_c$, the volume of the solution space shrinks to zero, indicating more patterns cannot be typically stored for {\it any} choice of perceptron weights. For continuous spherical weights, a RS calculation gives $\alpha_c=2$ and $\lim_{\alpha \to 2^-} S(\alpha) = -\infty$. For binary weights instead $\alpha_c\approx  0.83$, as can be obtained from the condition $S(\alpha_c)=0$. See  \cite{Engel2001} for an extensive discussion of the storage problem.

Some generalizations of the Hopfield model achieve super-extensive capacity: they are able to store a number of patterns polynomial \cite{Gardner1987multicon, Krotov2016, albanese_replica_2022} or even exponential \cite{Demircigil2017} in the size $N$ of the system. That is also true when continuous variables and memories are involved, as in the case of the modern Hopfield network of Ref. \cite{ramsauer2021hopfield}, a model linked to the wildly popular transformer architecture for deep learning \cite{vaswani2017attention}).

\subsection{Teacher-student scenarios}
From the theoretical analysis of the storage limits of neural networks, we move to learning problems, where the main interest is in characterizing the behavior of the generalization error. The theoretical analysis requires the statistical definition of: 1) a data generating process; 2) the model whose parameters have to be learned from the data. This framework is referred as the \emph{teacher-student model} in the literature. In the simplest scenario, both the teacher and the student are perceptrons, characterized by weight vectors $\bw^*$ and $\bw$ respectively. In an ideal Bayesian framework, the learner has access to the probability distribution  from which the teacher weights $\bw^*$ are generated and to the likelihood of producing a certain label $y$ given $\bw^*$ and an input $\bx$. The data generating process is the following:

\begin{align}
    \bw^* &\sim P_W, \\
    \bx^\mu &\sim P_X, \\
    y^\mu &\sim P\left(y\ \bigg|\   \frac{1}{\sqrt{\dimx}}\sum_{i} w_i^* x^\mu_i\right)  \qquad \mu =1,\dots,P.
\end{align}

In this Bayesian setting, the associated free energy is the log-normalization factor of the posterior distribution of the weights:
\begin{equation}
\phi = \lim_{\dimx\to\infty} \frac{1}{\dimx} \mathbb{E}_\dataset \log \int \dd\bw\  P(\bw) P(\dataset\, | \, \bw).
\end{equation}
This can be computed using the replica 
method. The calculation involves the introduction of a $n\times n$ overlap matrix $q_{ab}$ for the student and of a student-teacher overlap vector $r_a$:
\begin{align}
    q_{ab}=\frac{1}{\dimx}\sum_i w_i^a\, w_i^b; \qquad r_{a}=\frac{1}{\dimx}\sum_i w_i^a\, w^*_i. 
\end{align}
One then proceeds with a replica symmetry ansatz for these order parameters and sends the number of replicas $n$ to 0 as usual. In this optimal Bayesian setting,
where the student is statistically matched to the teacher, the Replica Symmetric ansatz is the correct one, thanks to the Nishimori condition \cite{nishimori_exact_1980,iba_nishimori_1999, Zdeborova2016}.
The free energy is simply expressed in terms of a few scalar integrals and 
obtained through saddle point evaluation of the order parameters. 

The expected generalization error is defined as the expectation (over the realizations of the dataset, the test example $(\bx, y)$, 
and the prediction $\hat{y}(\bx)$ from the model) of a cost function $c(\hat{y},y)$ comparing the true target and the prediction: 
\begin{equation}
    \mathcal{E}_{gen} = \mathbb{E}_\dataset\, \mathbb{E}_{\bx,y}\, \mathbb{E}_{\hat{y}\,|\,\bx,\dataset} \  c(\hat{y},y)
\end{equation}
In classification tasks, the cost function is typically the 0-1 valued error-counting function, while in regression instead it is the mean square error. 
Crucially, the expected generalization error for the Bayesian prediction $P(\hat{y} | \bx, \dataset) = \int \dd\bw\, P(\hat{y} |\bw) P(\bw|\dataset)$ can be simply expressed in terms of the saddle point order parameters. The whole RS replica picture for the Bayesian-optimal perceptron model has been rigorously established in Ref. \cite{Barbier2017a}.

In Section \ref{sec:overparam} we discuss the generalization of this approach to some simple models of two-layers multi-layer perceptrons \cite{Monasson1995, Schwarze1993} and other simple models. Reaching out to more complex teacher and student architectures, e.g. deeper and with proportional widths for all layers, is a major challenge for the statistical physics analysis of neural networks.

\subsection{Message passing algorithms}
\subsubsection{Belief Propagation equations}
The Belief Propagation (BP) algorithm \cite{Yedidia2002, Mezard2009} is a message passing algorithm for computing marginal and free energies in sparse graphical models. Prominent applications are in coding 
\cite{kabashima_belief_1998} and combinatorial optimization \cite{Krzakala2007} among others.
BP has been used to efficiently solve the problem of training a perceptron with binary weights \cite{Braunstein2006} where the iterations of the node-to-factor and factor-to-node messages take the form:

\begin{align}
m_{i\to \mu}^{(t+1)} &= \tanh\bigg(\sum_{\nu \setminus\mu} \xi^\nu_i\,\hat{m}_{\nu\to i}^{(t)}\bigg)    \\
\hat{m}_{\mu\to i}^{(t+1)} &= f\bigg(\sum_{j\setminus i} \xi^\mu_j\, m^{(t+1)}_{j\to\mu},\sum_{j} \big(m^{(t+1)}_{j\to\mu}\big)^2\bigg)     
\end{align}
with $f$ a simple function.
With respect to the standard BP prescription, the equations here have been simplified exploiting the central limit theorem since the factor graph is dense. This approximation goes under the name of relaxed Belief Propagation \cite{mezard_mean-field_2017}. Fixed points of the algorithm give the estimated marginals and can be used to make a Bayesian prediction for a given input. In order to produce a single binary configuration instead, one has to apply a decimation or reinforcement heuristic \cite{Braunstein2006} on top of BP.

\subsubsection{Approximate Message Passing and State Evolution}

It turns out that on dense graphical models such as the perceptron and under certain statistical assumptions on the disorder, the BP equations can be further simplified in what is known as the Approximate Message Passing algorithm (AMP). AMP was first proposed in a seminal paper by Donoho and Montanari \cite{Donoho2009} building on the rigorous analysis of Bolthausen of the TAP equations for the SK model \cite{bolthausen2014iterative}. Compared to BP, AMP has lower memory complexity since it involves the computation of only node-related quantities instead of edge ones. Moreover, in the high-dimension limit, the statistics of the AMP messages can be rigorously tracked by a dynamical system involving only a few scalar quantities, known as State Evolution (SE). Remarkably, SE involves the same quantities appearing in the RS replica calculation, and its fixed points correspond to the stationary points of the RS replica free energy  \cite{Barbier2017a}. 
Due to this connection, AMP has also been used as a proof technique for replica results in convex models \cite{loureiro_fluctuations_2022}.  See also \cite{Advani2016-te,Advani2016-du} for both replica and AMP perspectives on deriving optimal loss functions and regularizers for high dimensional regression. 

While Ref. \cite{Donoho2006} analyzed linear models, AMP was later extended to generalized linear models (where it is called GAMP) \cite{Rangan2011}, and committee machines with few hidden nodes (see Section \ref{sec:overparam}).

In inference settings, AMP has been applied to multi-layer architectures with an extensive number of hidden nodes but fixed weights \cite{Manoel2017b, Fletcher2018a}. The deep learning setting is much more challenging though, and there have been only limited attempts so far \cite{lucibello_deep_2022}.
Being able to scale message passing to deep learning scenarios would allow to perform approximate Bayesian estimation, train discrete weights for computational efficiency and energy saving, have analytically trackable algorithms, and perform cheap hyperparameter selection. 

\subsection{The geometry of the solution space}

Learning in Neural Networks (NNs) is in principle a difficult computational task: a non-convex optimization problem on a huge number of parameters. However, the problem seems to be relatively easy to solve, as even simple gradient-based algorithms convergence to solutions with good generalization capabilities. NNs models are evolving rapidly through a collective effort shared across many labs. It is thus difficult to define a unifying theoretical framework, and current NNs are in a sense similar to complex, highly evolved natural systems. 

A major question in deep learning concerns understanding the non-convex geometry of the error landscape as a function of the parameters, and how this geometry might facilitate gradient based learning.  Motivated by the geometry of random Gaussian landscapes \cite{bray2007statistics,fyodorov2007replica} (derived via replica theoretic methods), Ref. \cite{Dauphin2014-lk} numerically explored the statistics of extrema of the error landscape of deep and recurrent networks, finding that higher error extrema were typically higher index saddle points, not local minima.  Thus high error local minima need not confound deep learning, though low index saddle points might, and \cite{Dauphin2014-lk} developed an algorithm to rapidly escape these saddle points. 
The authors of \cite{baity2018comparing} undertook a careful comparison of the dynamics of stochastic gradient descent on neural networks versus the p-spin spherical spin glass energy function, finding  interesting ageing phenomena indicative of the prevalence of more flat directions as once descends the training error.  Ref. \cite{geiger2018jamming} found an interesting analogy between jamming and the error landscape of deep networks with a hinge loss, building on a prior analogy for the perceptron \cite{Franz2016-to}.  As the network size transitions from overparameterized (with many weight configurations at zero error) to underparameterized (with many isolated minima), the error landscape undergoes a jamming transition.  Finally Ref. \cite{Maillard2020-bn} provided another interesting exploration by extending the Kac-Rice method to count critical points in generalized linear models.  

A complementary view comes from some recent works that focus on the role played by  rare attractive minima. At least for  non-convex shallow neural networks classifying random patterns, it is possible to derive a theoretical description of the geometry of the  zero training error configurations (so-called "solutions"). The most immediate result is that the solutions that dominate the zero-temperature Gibbs measure of the error loss (i.e., the most numerous) do not match those found by the efficient learning algorithms. Numerical evidence suggests that in fact the solutions found by the algorithms belong to particularly entropic regions, i.e., with a high density of other nearby solutions \cite{PERCPRL15, UNREASONABLE16}. These types of solutions are often referred to as flat minima in the machine learning literature.

In deep learning settings,  numerical results consistently show that flatness of a minimizer positively correlates with generalization ability (see e.g. \cite{jiang_fantastic_2020}). Different algorithms explicitly targeting flat minima have been proposed \cite{Chaudhari2016a, pittorino_entropic_2021, foret2021sharpnessaware}.

\begin{figure}[t]
    \centering
    \includegraphics[width=\textwidth]{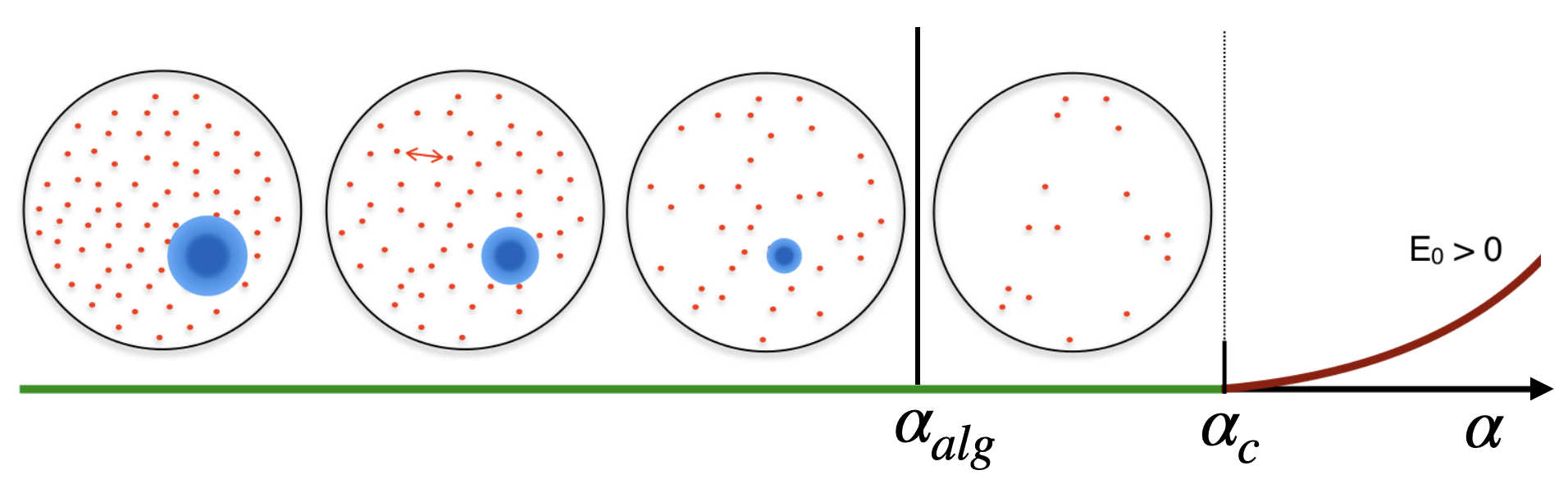}
    \caption{Geometrical transitions in the solution space for the storage problem of the binary perceptron at increasing load $\alpha=\ndata/\dimx$. Finding a solution in the connected cluster is easy, while finding a solution after the cluster disappears at $\alpha=\alpha_{alg}$ becomes computationally hard since all of them are isolated. Above $\alpha=\alpha_{c}$ no zero energy configurations exist.}
    \label{fig:geometry}
\end{figure}

The analytical study of flat minima and their generalization properties can be done by resorting to a large deviation technique introduced in Ref. \cite{PERCPRL15} and based on the Franz-Parisi potential \cite{franz_recipes_1995}. In order to introduce this framework, name Local Entropy (LE), it is useful to consider the simplest model displaying a rich geometry of the solution space, i.e. the binary perceptron. The LE framework can be applied to more complex architectures and continuous variables as well.
The local entropy function is defined as the (normalized) logarithm of the number of solutions $\bw^{\prime}$
at some  intensive distance $d$ from a reference solution $\bw$:

\begin{equation}
S_{LE}\left(d, \bw \right)=\frac{1}{N}\ln\, \sum_{\bw^\prime}
\, \mathbb{X}\left(\bw^{\prime}\right)\delta\big( d_H (\bw^{\prime},\bw) - d N\big)
\label{eq:energy_LE}
\end{equation}
where $d_H$ is the Hamming distance and $\mathbb{X}$ is the data-dependent indicator function for the solutions. The analysis of $S\left(d, \bw \right)$ for uniformly sampled solutions to the training set reveals that \emph{typical solutions are isolated} \cite{Huang2013}, meaning that no near solutions exist at small $d$.  This is the analogous of sharp minima in continuous networks.
It turns out that by sampling solutions according to the LE itself, that is from 

\begin{equation}
P_{\mathrm{LE}}\left(d, \bw\right)\propto e^{y N  {S}_{LE}\left(d,\bw\right)},
\label{eq:Gibbs_LE}
\end{equation}
it is possible to uncover the existence of \emph{rare} (according to the flat measure) regions containing an exponential number of solutions (a large volume in the case of continuous variables). By analogy, these are the flat minima found in continuous and deep architectures.
In the last equation $y$ has the role of an inverse temperature conjugated to the local entropy. For large values of $y$ the probability  focuses on the $\bw$ which are surrounded by an exponential number of solutions at distance $d$, therefore suppressing isolated solutions when $d$ is small enough. Fig. \ref{fig:geometry} displays the coexistence of isolated and dense solutions up to a certain value of $\alpha$ when the dense region (in blue) disappears. In Ref. \cite{PERCPRL15} it is analytically shown that dense solutions generalize better than isolated ones. It must be also noted that isolated solutions cannot be algorithmically accessed by polynomial algorithms.
Rigorous results on the geometry of the symmetric variant of the binary perceptron have been obtained  in Ref. \cite{abbe2022binary}. 

Adapting the LE definition to generic architectures and continuous weights, one can try direct optimization of $S_{LE}$ instead of the original loss.  Estimation and optimization can be done within a double-loop algorithm known as Entropy-SGD \cite{Chaudhari2016a, pittorino_entropic_2021}. Second order approximation of $S_{LE}$ leads to the Sharpness Aware Minimization algorithm of Ref. \cite{foret2021sharpnessaware}.

Other algorithmic approaches for the generic deep learning setting are obtained as follows \cite{UNREASONABLE16}. Starting from  Eq. \eqref{eq:Gibbs_LE}, relax the constraint on the distance with a Lagrange multiplier $\gamma$, replace the indicator function in Eq. \eqref{eq:energy_LE} with a Boltzmann weight involving a loss ${\cal L}(\bw^\prime)$, 
\emph{take integer} $y$, and unfold the exponential in Eq. \eqref{eq:Gibbs_LE} by introducing $y$ replicas of the original system. 
We end up with a replicated system with an effective energy  

\begin{equation}
{\cal L}_{\mathrm{R}}\left(\bw,\{\bw^{\prime a}\}_a\right)=\sum_{a=1}^{y}{\cal L}\left(\bw^{\prime a}\right)+\gamma\sum_{a=1}^{y}d\left(\bw,\bw^{\prime a}\right).\label{eq:L_R}
\end{equation}

We thus have a central replica interacting with $y$ peripheral ones that also resent from the data-dependent loss function. As discussed in ref.~\cite{UNREASONABLE16, baldassi2016local, baldassi2020shaping}, several algorithmic schemes can be derived in a straightforward way  by  optimizing the replicated loss function ${\cal L}_{\mathrm{R}}$, for example with gradient-based algorithms, or by sampling the space of solutions with a Markovian process or with Belief Propagation equations.
One of these, the flat-minima-seeking algorithm known as rSGD, has been applied to deep neural architectures in Ref. \cite{pittorino_entropic_2021}.

\subsection{Learning dynamics}
\label{sec:perceptron:online}
With neural network models, the dynamics of gradient descent takes place in a high-dimensional non-convex landscape, therefore its theoretical description is highly non-trivial. For the perceptron and in the continuous time limit, dynamical mean-field theory has been used to provided such description in terms of low-dimensional integro-differential equations involving two-times correlations functions and responses \cite{agoritsas_out--equilibrium_2018}. Such description has been extended to stochastic gradient descent as well \cite{mignacco_dynamical_2020}. Numerical solution of the equations is particularly challenging, and the framework has not been extended to deeper architectures so far.  

In the simpler online setting, where only a single pass over the data points is allowed, a much simpler set of ODE can be derived. For the committee machine architecture detailed in Section \ref{sec:overparam} the quantities to be evolved are the student-student and teacher-student overlaps among the hidden perceptrons, $Q_{k,k'}$ and $R_{k,k^*}$ respectively \cite{Saad1995, Saad1995a, Saad1999a, Goldt2019}. In terms of these overlaps, the evolution of the generalization error can be described at each time step.
In the same setting, but assuming infinitely wide networks with finite input size, the authors of Ref.~\cite{mei_mean_2018} obtained a mean-field description in terms of a PDE characterizing a diffusion process for the perceptrons' weights.

\section{Studying over-parametrized models}
\label{sec:overparam}
The repeated breakthroughs of deep learning starting from the 2010's propelled neural networks to the forefront of machine learning \cite{LeCun2015a}. Moreover these neural networks increased rapidly in size; one of the first convolutional networks for document recognition, LeNet-5, had around $61,000$ trainable parameters \cite{LeCun1998}, while today's GPT-3 has around $175$ billion \cite{Brown2020}. This proliferation of trainable parameters leads to highly flexible models, which begs the question of why they don't overfit to their training data and why they can still generalize well to new input examples.  Explaining their success in this so-called \emph{over-parametrized} regime constitutes a key theoretical question in machine learning, as it seemingly violates the classical picture of the bias-variance trade-off \cite{Spigler2018,Belkin2018}. In this section, we describe how over-parametrization has been tackled by the statistical mechanics approach.

\begin{figure}[t]
    \centering
    \includegraphics[width=\textwidth]{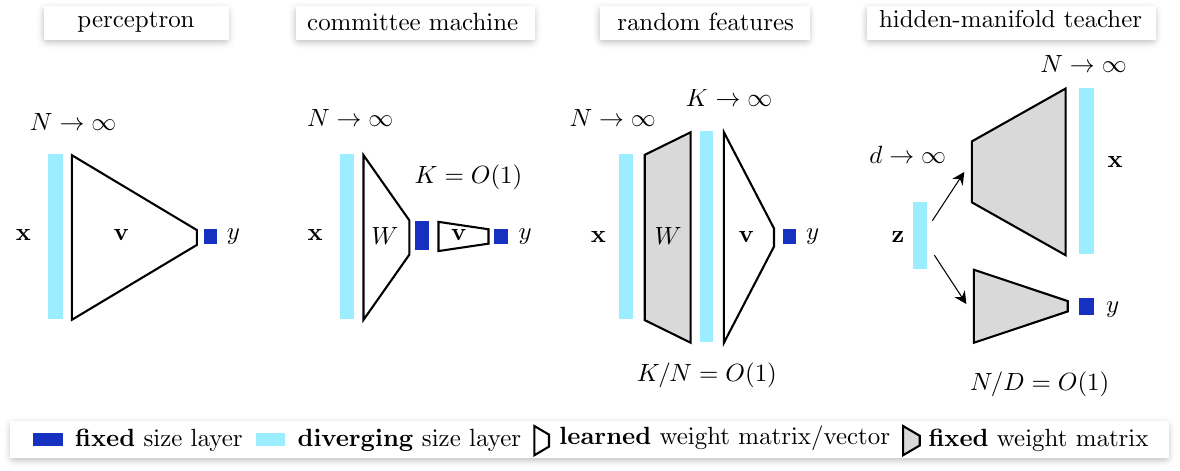}
    \caption{Scaling comparison of models' dimensions in the thermodynamic limit.}
    \label{fig:scalings}
\end{figure}

\subsection{Committee machines}

Gardner's program for the perceptron, described above, was soon extended to multi-layer networks. The first analyzed model consisted of summing the outputs of multiple perceptrons with non-overlapping inputs \cite{Mato1992}. This tree-like architecture was called \emph{committee machine}. Soon after, this analysis was extended to full connectivity to the input \cite{Schwarze1993, Schwarze1993a}. These architectures are cases of a generic one-hidden layer neural-net, which passes a linear combination of the outputs of $\nhid$ perceptrons, each with non-linearity $\sigma(\cdot)$, through an output activation function $f(\cdot)$, yielding a final output $y$ given by 
\begin{align}
    \label{eq:committee-in-out}
    y = f \left(\sum_{k=1}^K v_k \; \sigma \, \left( \sum_{i=1}^{\dimx} W_{ki} \, x_i\right)\right).
\end{align}
The input to hidden layers weights are parameterized by a matrix $W \in \R^{\nhid \times \dimx}$ and the hidden to output weights are parameterized by a vector $\bv \in \R^\nhid$. $\nhid$ denotes the number of \emph{hidden} neurons.

Remarkably, any smooth function on $\R^\dimx$ can be arbitrarily well approximated by  \eqref{eq:committee-in-out} with a finite, but possibly large, $\nhid$ \cite{Hornik1991}. This universal approximation theorem motivates the one-hidden layer neural network as a simple yet far-reaching learning model to be studied. Also, these one hidden layer networks allow a study of over-parametrization by considering large student networks with $K$ hidden units learning from data generated by small teacher networks with $M < K$ hidden units. 

\paragraph{Scaling and specialization transition --} The statistical mechanics analysis of committee machines in the teacher-student scenario mirrors that of the perceptron. Given a dataset $\dataset = \{\bx_\mu, y_\mu\}_{\mu=1}^\ndata$ generated by a teacher committee machine with $M$ hidden units of the form 
\begin{align}
    \label{eq:committee-in-out-teacher}
    y^\mu = f^* \left(\sum_{k=1}^M v^*_k \, \sigma \, \left( \sum_{i=1}^{\dimx} W^*_{ki} \, x^\mu_i\right)\right),
\end{align}
one can derive learning curves in the usual high dimensional limit $\ndata,\dimx \to \infty$ with $\alpha = \ndata/\dimx$ held to be $O(1)$. Also the number of hidden units in the teacher ($M$) and student ($K$) are both kept $O(1)$ (see Figure \ref{fig:scalings}). The order parameters are the teacher-student and student-student overlaps, as for the perceptron, except now they become matrices for the input layer:
\begin{align}
    R = \frac{W W^{*,\top}}{\dimx}\in \R^{K \times M}, Q = \frac{W W^\top}{\dimx} \in \R^{K \times K}. 
\end{align}
For simplicity we assume the teacher output $\bv^* \in \R^\nhid$ to be the all 1 vector. 

\begin{figure}[t]
    \centering
    \includegraphics[width=0.7\textwidth,trim={0 2cm 1cm 0},clip]{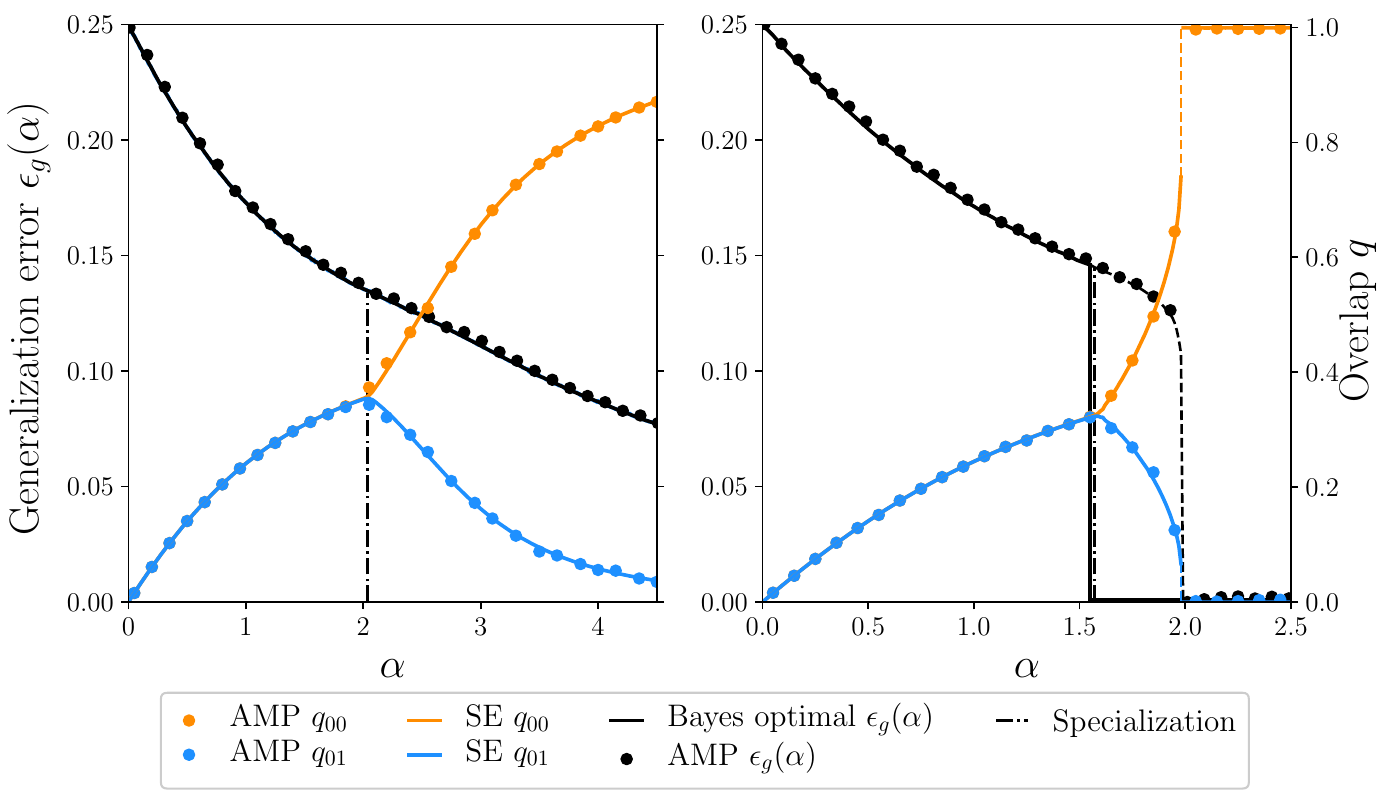}
    \includegraphics[width=0.2\textwidth]{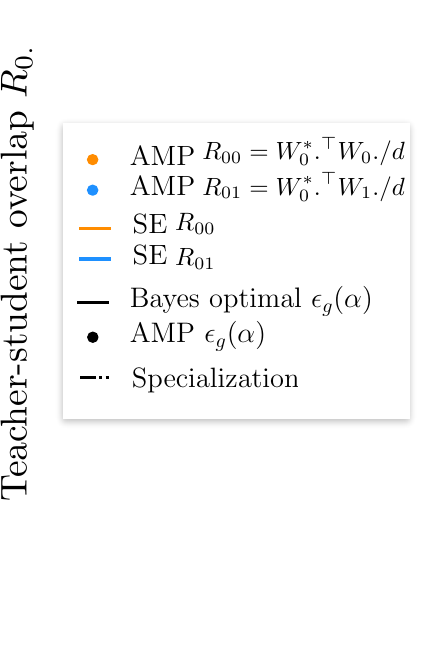}
    \caption{Adapted from \cite{Aubin2018}. Overlap parameters and generalization for matched teacher-student committee machines with $M=\nhid=2$ hidden units as a function of the ratio $\alpha=\ndata/\dimx$ between sample size and input dimension, with first layer weights either Gaussian (left) or binary (right). As $\alpha$ grows beyond a critical value, the overlaps account for the specialization of the student hidden units to the teacher's.}
    \label{fig:committee-K2}
\end{figure}

A remarkable phenomenology was identified using annealed and quenched replica computations \cite{Schwarze1993, Schwarze1993a} as well as an online learning analysis \cite{Saad1995, Saad1995a, Biehl1995}. Considering  matched teacher and students ($M=K$) with fixed student output weights $\bv = \bv^*$, the Bayes optimal student weights were shown to specialize to the teacher weights only if a critical amount of data was available (i.e. $\alpha > \alpha_c$, see Figure \ref{fig:committee-K2}). This specialization transition was derived rigorously recently in \cite{Aubin2018}, through AMP and SE. Non-linear hidden units are necessary for specialization to set in, and in scarce data regimes ($\alpha < \alpha_c$) committee machines act no differently than linear models such as perceptrons. 

\begin{figure}[t]
    \centering
    \includegraphics[width=\textwidth]{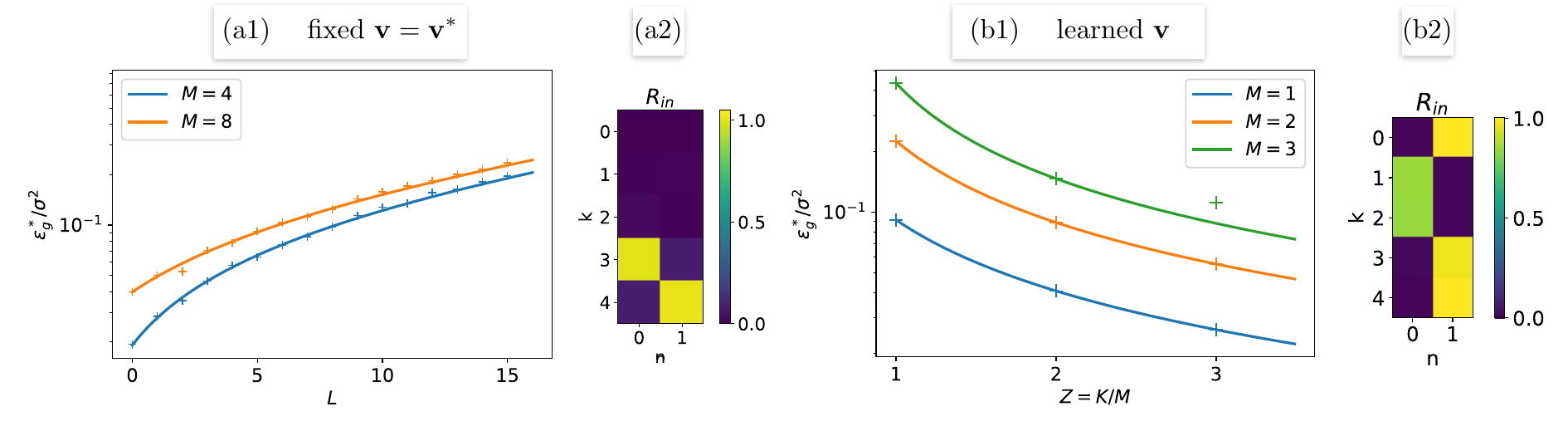}
    \caption{Adapted from \cite{Goldt2019}. (a1) and (b1) Generalization error as a function of the over-parametrization for sigmoidal committee machines with $\nhid$ hidden units learning from data generated with teachers with $M$ hidden units ($L = \nhid - M$). Generalization deteriorates with over-parametrization when the output weight vector $\bv$ is fixed (a1) and improves when $\bv$ is learned (b1). This behavior is traced back to the specialization of multiple hidden units for each of the teacher's hidden unit in this second case only as testified by the reported teacher-student overlap matrices (a2) and (b2) for $M=2$ and $\nhid=5$. }
    \label{fig:committee-denoise}
\end{figure}

\paragraph{Denoising solution --}
The online analysis of the specialization transition was also made rigorous recently by \cite{Goldt2019}. In this work, Goldt and co-authors further extended the analysis to include learnable student output weights $\bv$ and uncovered one mechanism of improved generalization by over-parametrization. Focusing on committee machines with sigmoidal non-linearities, they showed that the generalization error increases (decreases) with overparameterization $L = \nhid - M$ if the output layer weight $\bv$ is fixed (learned) (see Figure \ref{fig:committee-denoise}). The natural order parameters, or overlap matrices $R$ and $Q$, once again, elucidate the mechanism behind this phenomenology. With learnable output weights, several of the $\nhid$ hidden units of the student can (potentially weakly) specialize to one of the $M$ teacher hidden units, and the student output weights can learn to combine, or denoise, these student hidden-unit outputs to mimic the overall teacher output. This \emph{denoising} solution allows the student with more hidden units to build a larger ensemble of regressors of each of the teacher's hidden units, thereby improving its accuracy with overparameterization. Conversely, when the student output weight $\bv$ is fixed, only $M$ among the $\nhid$ student hidden units can effectively participate in the student committee, each specializing to one of the teacher's units. The remaining $L = \nhid-M$ units only contribute noise, so overparameterization hurts generalization.

\subsection{Kernel-like learning}
The analysis of committee machines revealed a fundamental mechanism of generalization through over-parametrization, yet, these models operate in an atypical regime where the input dimension diverges while the number of hidden units is fixed.
Operating in a different scaling, kernel methods form another class of models that can be analyzed in depth (see Chap 16 \cite{Shalev-Shwartz2014}). Kernel-based learners are non-linear in the inputs but linear in the parameters. While deep neural networks are in essence more expressive than kernel methods\footnote{Kernel models are not universal approximators.}, they sometimes behave not much differently from kernels, in particular in cases where they have infinitely wide hidden layers \cite{Jacot2018} (though see \cite{Fort2020-wn} for empirical differences between kernel learners and deep networks at more natural widths). Moreover kernel models already exhibit key features to be understood in deep learning. Notably, the generalization performance of kernel learners undergoes a \emph{double descent} \cite{Advani2020,Spigler2018, Belkin2018}: after an overfitting peak predicted by the classical bias-variance compromise, the generalization improves continuously as the number of parameters increases. Therefore, kernel methods also enjoy good generalization with over-parametrization. 

\begin{figure}[t]
    \centering
    \includegraphics[width=\textwidth]{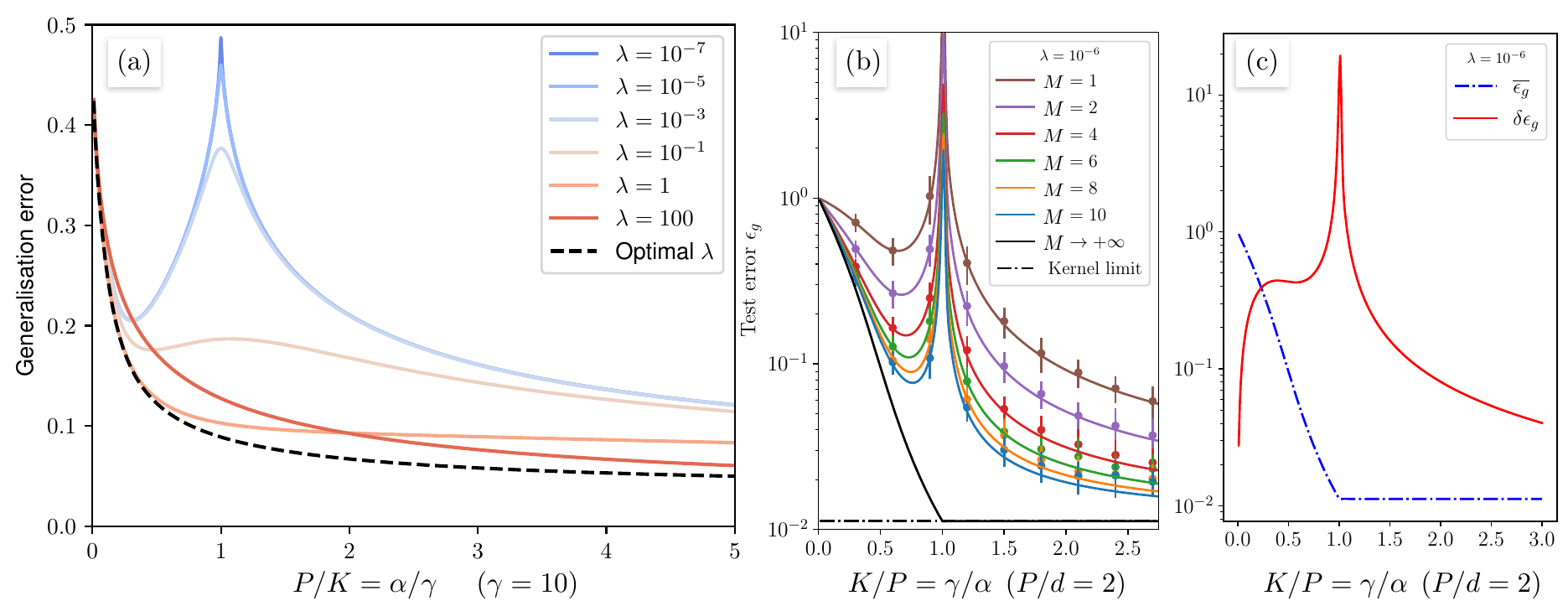}
    \caption{The double descent analyzed in random-feature models through replica computation of the generalization error. (a) As an $\ell_2$ regularization of strength $\lambda$ is adjusted, the overfitting peak at $\ndata/\nhid=1$ vanishes. (b) Similarly, averaging the prediction over an ensemble of $M$ models mitigates the double descent pointing to the origin of the overfitting in exploding variance, as the computed decomposition shows in (c).  Adapted from \cite{Gerace2020} (a) and \cite{Loureiro2022} (b) and (c).}
    \label{fig:double-descent}
\end{figure}

\paragraph{Random features and Gaussian equivalence --}
The random feature model \cite{Rahimi2017}, is a simple kernel-like model that is closely related to the one-hidden layer model defined above in \eqref{eq:committee-in-out}. It only differs by fixing first-layer weights $W$ to random values from a given distribution, such as a Gaussian or select Fourier modes at random frequencies. The only learnable parameters are in the output weight $\bv \in \R^\nhid$, and the flexibility of the model can be controlled by the number of hidden units $\nhid$ (see Figure \ref{fig:scalings}). 
In the scaling limit where $\nhid$ and the sample size $\ndata$ scale linearly with the dimension of input $\dimx$, the generalization error can be characterized asymptotically through different methods. With the additional assumption that the input data is Gaussian distributed and the teacher input-output rule is a perceptron: $y^\mu = f^* \left({\bv^*}^\top \bx^\mu \right)$, learning curves were derived using random matrix theory \cite{Mei2021},  replica computations \cite{Gerace2020,DAscoli2021} and Gaussian convex inequalities \cite{Dhifallah2020}. An important insight common to these analyses is the equivalence of the non-linear features with a Gaussian model with matching moments.  That is the generalization error averaged over the teacher data distribution
\begin{align}
    E_{g} = \E_{\bx, y}\left[ \left(f \left(\bv^\top \sigma \, \left( W \, \bx\right)\right) -  f^* \left({\bv^*}^\top \bx^\mu \right) \right)^2 \right], 
\end{align}
concentrates in the thermodynamic limit around a function of the ratios $\alpha = \ndata/\dimx$ and $\gamma = \nhid/\dimx$ given by
\begin{align}
    \mathcal{E}_{g}(\alpha, \gamma) = \lim_{\dimx \rightarrow \infty}  E_{g} =\int_{\R^2}  \left(f(\lambda) - f^*(\nu)\right)^2 \N(\lambda, \gamma \, ; \, 0, \Sigma_{\alpha, \gamma})d \lambda d \nu\,,
\end{align}
where the covariance of the jointly Gaussian scalars $\lambda$ and $\nu$ are themselves function of the overlap order parameters of the problem and constants accounting for the choice of non-linearity $\sigma$ and input distribution. This so-called \emph{Gaussian equivalence principle} \cite{Goldt2020} was first described for square losses using rigorous random matrix theories \cite{Pennington2017, Mei2021} and later extended to generic convex losses \cite{Goldt2021, Hu2020}. This formulation finally allows to predict theoretically the origin of the overfitting peak in the double descent. At the interpolation threshold, where the learning model has just enough parameters to perfectly fit the training data, the variance of the learned predictor explodes \cite{DAscoli2020, Mei2021, Loureiro2022}. Beyond the interpolation threshold, over-parametrized models feature many minimizers of the training loss, yet implicit or explicit regularization of the training typically selects a minimizer with good generalization (see Figure \ref{fig:double-descent}).

\paragraph{Universality --} 
Beyond random features in a teacher-student scenario, kernel-learning has been analyzed in great generality using statistical mechanics tools. Loureiro and collaborators first replaced shallow random maps with learned deep feature maps \cite{Loureiro2021}, and then formulated a universal theory for any generalized linear model trained through the empirical risk minimization of a convex loss \cite{Loureiro2022}. Interestingly, the latter result is mathematically rigorously proven through a novel proof technique relying on AMP \cite{Gerbelot2021}.

Focusing instead on a generic formulation of kernel regression, Canatar and collaborators developed a widely applicable theory of learning encompassing arbitrary training data \cite{Canatar2020}. A replica computation yields the generalization error as a function of the kernel spectral decomposition, the training data distribution and the sample size. This analysis highlights the impact of the data distribution through the eigenmodes of the kernel and through the so-called \emph{alignment} between the data and the task to be learned. Along these lines, one can show in simple linear settings that learning curves with respect to the amount of data can exhibit an arbitrary number of multiple descent peaks equal to the number of scales in the data \cite{Mel2021-dr}. This further highlights the importance of understanding how non-random structure in data impacts learning properties of neural networks. 

\subsection{Studying the impact of structure in data}
The question of how and when structure in data drives generalization in neural networks is fundamental in learning theory. The sequence of above works can be viewed as the successive addition of structure in data, starting from random uncorrelated inputs and output labels in the storage problem,  to random i.i.d inputs that are correlated to their corresponding outputs through the introduction of a teacher.  More recent work has examined the important generalization of going beyond i.i.d. random inputs.  

A first notable step is the introduction of the \emph{hidden manifold} model assuming that the data and labels are generated from a low dimensional representation \cite{Goldt2019a} (see Figure \ref{fig:scalings}). The analytical description of learning in committee machines, relying on the Gaussian equivalence principle, shows that generalization is tied to the dimension of the underlying manifold. These results are consistent with the analyses of random-feature models trained on anisotropic Gaussian data combining a high-variance \emph{strong} subspace and a low-variance \emph{weak} subspace \cite{Ghorbani2020,DAscoli2021}. The concept of data-task alignment is here again identified as a determinant of generalisation: it ensures that the effective dimension of the problem is small and allows for good generalisation at small sample size.

Another model of structure in data, especially relevant for classification problems, is the Gaussian mixture model of inputs.  Limits of separability and generalization behavior can be derived for classification of Gaussian clusters with generalized linear models \cite{Loureiro2021a} and committee machines \cite{Refinetti2021}. In particular, the latter work demonstrates that some configurations of clusters unlearnable with random features (an instance of a generalized linear model) can be separated by committee machines. The 2-layer neural network learns appropriate features to allow for a proper task-data representation alignment, whereas generalized linear models are limited by their fixed feature-maps, which may not be appropriate for the task. 

\subsection{Non-convex overparametrized neural networks}

Finally, in ref. \cite{baldassi2022learning}  the computational consequences  of overparameterization in non-convex neural network models are analyzed. In the simple case of discrete binary weights, a non-convex classifier is connected to  a random features projection. The analysis shows that  as the number of connection weights increases multiple phase transitions happen in the zero error landscape. A first transition happens at the so-called interpolation point, when solutions begin to exist (perfect fitting becomes possible). A second transition occurs  with the discontinuous appearance of a different kind of “atypical” structures corresponding to  high local entropy regions with good generalization properties.

\section{Going deeper}
\label{sec:deep}

Much of the remarkable progress in artificial intelligence over the last decade has been driven by our ability to train very deep networks with many successive nonlinear layers. Formally, the simplest version of a feedforward neural network with $D$ layers is the multi-layer perceptron, defined through 

\begin{equation}
\x^{l} = \phi(\h^l) \qquad 
\h^l = \Wl{l} \, \x^{l-1} + 
\bi^l \quad \text{for} \, \, \, l=1,\dots,D. 
\label{eq:netdynam}
\end{equation}
Here, $\x^{0}\in \mathbb{R}^{N_0}$ is the input, which propagates through the network to generate a sequence of activity vectors $\x^{l}\in \mathbb{R}^{N_l}$ in layer $l$ with $N_l$ neurons. $\Wl{l}$ is an $N_{l} \times N_{l-1}$ weight matrix connecting neurons in layer $l-1$ to layer $l$, $\bi^l$ is a vector of biases to neurons in layer $l$, $\h^l$ is the pattern of inputs to neurons at layer $l$, and 
$\phi$ is a single neuron scalar nonlinearity that acts component-wise to transform inputs $\h^l$ to activities $\x^l$.  The final output of the network is $\mb{y} = \x^D(\x^0,\mb{w})$ where $\mb{w}$ collectively denotes all $N$ neural network parameters $\{\Wl{l} , \bi^l\}_{l=1}^D$.  

The theoretical analysis of such deep networks in full generality raises several difficult challenges which remain open to this day \cite{Bahri2020-mi,Gabrie2019,Maillard2021}. For example, how can we describe their learning dynamics? What can deep networks express that their shallow counterparts cannot? How should we initialize the weights to optimize their learning dynamics? What does their training error landscape look like? How can we use them for generative modeling? How does their performance scale with network size? In the following we review some recent progress on these questions that involves ideas from both equilibrium and non-equilibrium statistical mechanics, nonlinear dynamical systems, random matrix theory, and free probability. 

\subsection{Exact learning dynamics for deep linear networks}

A gold standard of understanding deep learning would be an exact solution to the learning dynamics of both training and generalization error as a function of training time, for arbitrarily deep networks and for arbitrarily structured data.  While this is challenging for networks of the form in \eqref{eq:netdynam}, remarkably it is possible in the case of linear networks with $\phi(x)=x$ and squared loss \cite{Saxe2013,Lampinen2018-sl}. Despite the linearity of the network, the learning dynamics is highly nonlinear. Exact solutions to these dynamics reveal that deep linear networks learn by successively approximating the singular value decomposition (SVD) of the input-output correlation matrix of the training data mode by mode.  Each mode is learned on a time scale inversely related to its singular value. 

Intriguingly, while deep nonlinear networks have long been used in psychology to model the developmental dynamics of semantic cognition infants \cite{Rogers2004-px}, a recent analysis  \cite{Saxe2019-eg} showed that much of this developmental learning dynamics could be qualitatively captured by deep linear networks, thereby accounting for a diversity of phenomena, including progressive differentiation of semantics, semantic illusions, item typicality, category coherence, dynamic patterns of inductive projection, and the conservation of semantic similarity across species.  

Additionally, recent advances in self-supervised learning \cite{Chen2020-ib,Grill_undated-ho} yield new types of learning dynamics that can also be well modelled by deep linear networks \cite{Tian2020-bi,Tian2021-wg}, even in settings where the learning dynamics do not correspond to gradient descent on any function \cite{Grill_undated-ho}.  Moreover, these simpler learning models have prescriptive value: analytically derivable hyperparameter choices that work well for training deep linear models, also work well for their highly nonlinear counterparts \cite{Tian2021-wg}, thereby opening the door to the use of mathematical analysis to drive practical design decisions.

\subsection{Expressivity and signal propagation in random nets}

Even before one trains a network, one has to choose the initial weights and biases, for example $\{\Wl{l} , \bi^l\}_{l=1}^D$ in \eqref{eq:netdynam}, and the choice of such an initialization can have a dramatic practical impact on subsequent learning dynamics. A common choice is a zero mean i.i.d. Gaussian initialization with variance $\sigma^2_w/N_{l-1}$ for weights $\Wl{l}_{ij}$ and variance $\sigma^2_b$ for biases $b^l_i$. This relative scaling ensures weights and biases exert similar control over a neuron in layer $l$ as any previous layer width $N_{l-1}$ becomes large. 

In the limit of large $N_l$, \cite{Poole2016-fq} analyzed the propagation of signals through \eqref{eq:netdynam} via dynamic mean field theory and found an order to chaos phase transition in the $\sigma^2_w$ by $\sigma^2_b$ plane for sigmoidal nonlinearities $\phi$. In the ordered regime with small $\sigma^2_w$ relative to $\sigma^2_b$, the network contracts nearby inputs as they propagate forward through the network and backpropagated error gradients vanish exponentially in depth. In the chaotic phase with $\sigma^2_w$ large relative to $\sigma^2_b$, forward propagation chaotically amplifies and then folds small differences in inputs, leading to highly flexible and expressive input-output maps,  while backpropagated error gradients explode exponentially in depth.  Initializing networks near the edge of chaos in the $\sigma^2_w$ by $\sigma^2_b$ plane yields good, well-conditioned initializations. Indeed \cite{Schoenholz2017} found that the closer one initializes to the edge of chaos, the deeper a network one can train.   

Subsequent work \cite{Pennington2017-za,Pennington2018-fy} showed that one can go beyond i.i.d. Gaussian initializations to speed up learning especially in very deep or recurrent networks. They considered the Jacobian
\begin{equation}
\label{eqn:Jz}
{J} = \frac{\partial \mb{x}^D}{\partial \mb{x}^0} = \prod_{l=1}^D {D}^l {W}^l.
\end{equation}
Here ${D}^l$ is a diagonal matrix with entries $D^l_{ij} = \phi'(h^l_i) \, \delta_{ij}$ and $\mb{h}^l$ is defined in \eqref{eq:netdynam}.  $J$ measures the susceptibility of the network's output to small changes in the input. Related susceptibilities play a fundamental role in the backpropagation of error that guides gradient based learning. Initializing at the edge of chaos controls the mean squared singular value of $J$ to be $1$. But for very deep networks, $J$ could still become ill conditioned with the maximal singular value growing at a rate that is linear in the depth, even at the edge of chaos.  \cite{Pennington2017-za,Pennington2018-fy} controlled this growth by analytically computing the {\it entire} singular value spectrum of $J$ using free probability, exploiting the fact that $J$ is a product of random matrices, and then determined how and when we can ensure {\it dynamically isometric} initializations in which the {\it entire} singular value distribution  can be tightly concentrated around $1$. In particular, orthogonal weight matrices with sigmoidal nonlinearities achieve such dynamical isometry, which was shown to speed up training.  

Building on these works \cite{Xiao2018-ns} extended dynamically isometric initializations to convolutional neural networks, and showed how to train 10,000 layer models without any of the complex normalization tricks used in deep learning. Thus overall, studies of signal propagation and initialization in random deep networks provide an example of how statistical mechanics type analyses can guide engineering design decisions.

\subsection{Generative models via non-equilibrium dynamics}

A recent major advance in deep learning is the ability to generate remarkable, novel photorealistic images from text descriptions (see e.g. \cite{Rombach_undated-yk,saharia2022photorealistic,ramesh2022hierarchical}). Intriguingly one component of models that can do this well was inspired by ideas in non-equilibrium statistical mechanics \cite{Sohl-Dickstein2015-sb,ho2020denoising}.  In particular, being able to generate images requires modeling the probability distribution over natural images.  Equilibrium methods for modeling complex distributions involve creating a Markov chain that obeys detailed balance with respect to the distribution. However, if the distribution has multiple modes, such Markov chains can take very long to mix. 

\cite{Sohl-Dickstein2015-sb} instead suggested training a finite time nonequilibrium stochastic process to generate images starting from noise. The method of training involved taking natural images and allowing them to diffuse in a high-dimensional pixel space, thereby destroying their structure and converting them to noise. Then a neural network was trained to reverse the flow of time in this otherwise irreversible structure-destroying diffusion process. The result is then a trained neural network that can sample images by converting any white noise image into a particular naturalistic image through an approximate reversal of the diffusion process.  Thus this provides another example of how ideas in statistical mechanics can eventually lead to state of the art systems in artificial intelligence.

\subsection{Neural scaling laws governing deep learning}

Another major shift in deep learning over the last few years has been the immense scaling up of both dataset and model sizes. This has partially been motivated by empirically observed neural scaling laws \cite{Hestness2017-yq,Kaplan2020-ti,Henighan2020-jf,Gordon2021-az,Hernandez2021-ix,Zhai2021-dl,Hoffmann2022-gw} in many domains of machine learning, including vision, language, and speech, which demonstrate that test error often falls off as a power law with either the amount of training data, model size, or compute. Such power law scaling has motivated significant societal investments in data collection, compute, and associated energy consumption. However, obtaining a theoretical understanding of the origins of these scaling laws, and the dependence of their exponents on structure in data, model architecture, learning hyperparameters, etc... constitutes a major research question \cite{Bahri2021-yr}. 

Also interesting is whether power law neural scaling can be beaten. Recent work \cite{Sorscher2022-wo} explored scaling with respect to dataset size to analyze wether intelligent data pruning methods \cite{Paul2021-ci} that involve careful subselection of training data, could lead to more efficient scaling of error w.r.t. pruned dataset sizes.  Returning to the perceptron, \cite{Sorscher2022-wo} developed a replica calculation to compute the test error for non-Gaussian pruned data, and showed it is possible to beat power law scaling, and even approach exponential scaling, provided one has access to a good data pruning metric.  Then going beyond the perceptron \cite{Sorscher2022-wo} showed how to beat power law scaling in modern architectures like ResNets trained on modern datasets, including ImageNet.  

Overall, a deeper understanding of the origin of neural scaling laws, and the emergent capabilities that arise \cite{Wei2022-hw}, constitutes a major open question, which may benefit from the analysis of appropriate statistical mechanics models capturing the essence of these phenomena.  



\bibliographystyle{ws-book-har}
\bibliography{bibliography}

\end{document}